# Output and ($k_{Q_{clin},Q_{msr}}^{f_{clin},f_{msr}}$) correction factors measured and calculated in very small circular fields for microDiamond and EFD-3G detectors

Eyad Alhakeem [1, 2] and Sergei Zavgorodni [2,1]

[1] Department of Physics and Astronomy, University of Victoria, Victoria, British Columbia V8W 3P6, Canada
[2] Department of Medical Physics, British Columbia Cancer Agency–Vancouver Island Centre, Victoria, British Columbia V8R 6V5, Canada

Email: eyadali@uvic.ca

The purpose of this work was to obtain $k_{Q_{clin},Q_{msr}}^{f_{clin},f_{msr}}$ factors for microDiamond and EFD-3G detectors in very small (less than 5 mm) circular fields. We also investigated the impact of possible variations in microDiamond detector design schematics on the calculated $k_{Q_{clin},Q_{msr}}^{f_{clin},f_{msr}}$ factors. Output factors (*OF*'s) of 6 MV beams from TrueBeam linac collimated with 1.27-40 mm diameter cones were measured with EBT3 films, microDiamond and EFD-3G detectors as well as calculated (in water) using Monte Carlo (MC) methods. Based on EBT3 measurements and MC calculations $k_{Q_{clin},Q_{msr}}^{f_{clin},f_{msr}}$ factors were derived for these detectors. MC calculations were performed for microDiamond detector in parallel and perpendicular orientations relative to the beam axis. Furthermore, $k_{Q_{clin},Q_{msr}}^{f_{clin},f_{msr}}$ factors were calculated for two microDiamond detector models, differing by the presence or absence of metallic pins. The measured *OF*s agreed within 2.4% for fields ≥10 mm. For the cones of 1.27, 2.46, and 3.77 mm maximum differences were 17.9%, 1.8% and 9.0%, respectively. MC calculated output factors in water agreed with those obtained using EBT3 film within 2.2% for all fields. MC calculated $k_{Q_{clin},Q_{msr}}^{f_{clin},f_{msr}}$ factors for microDiamond detector in fields ≥10 mm ranged within 0.975-1.020 for perpendicular and parallel orientations. MicroDiamond detector $k_{Q_{clin},Q_{msr}}^{f_{clin},f_{msr}}$ factors calculated for the 1.27, 2.46 and 3.77 mm fields were 1.974, 1.139 and 0.982 with detector in parallel orientation, and these factors were 1.150, 0.925 and 0.914 in perpendicular orientation. Including metallic pins in the microDiamond model had little effect on calculated $k_{Q_{clin},Q_{msr}}^{f_{clin},f_{msr}}$ factors. EBT3 and MC obtained $k_{Q_{clin},Q_{msr}}^{f_{clin},f_{msr}}$ factors agreed within 3.7% for fields of ≥3.77 mm and within 5.9% for smaller cones. Including metallic pins in the detector model had no effect on calculated $k_{Q_{clin},Q_{msr}}^{f_{clin},f_{msr}}$ factors.

## 1. INTRODUCTION

Small photon beams are often used in modern radiotherapy to treat brain tumors and functional disorders. Trigeminal neuralgia is a brain nerve disorder where treated volume can be as small as 2 mm and is typically irradiated with a single high dose fraction of radiation in the order of 80 Gy (Urgošik et al 1998). Also, in IMRT and VMAT delivery of stereotactic radiotherapy treatments of small tumors apertures of the size of 2 mm and less are not uncommon. Therefore, establishing accurate dosimetry in 1- 3 mm diameter fields becomes important for quality assurance protocols and safe delivery of such treatments.

Das *et al* (2008), summarized the challenges associated with small field dosimetry. These include lack of lateral charge equilibrium, source occlusions, and detector perturbations. In order to minimize possible dosimetric errors, the IAEA-AAPM TRS-483 (Palmans *et al* 2017) provided guidelines on small field dosimetry. This report defines small fields, provides recommendations on suitable detectors and good working practice for dosimetry in such conditions.

Scott *et al* (2008), investigated dosimetric properties of 5x5 mm$^2$-100×100 mm$^2$ photon fields. Output factors (*OF*) measured by all dosimeters were in agreement to within 1% for fields greater than 20×20 mm$^2$, and within 2.8% for the 15×15 mm$^2$ field. However, for the 5×5 mm$^2$ field, the maximum difference in measured *OF* was 8.5%.

Marsolat *et al* (2013), compared *OF* measurements made with a single crystal diamond dosimeter (SCDDo) against other small field detectors for 6 MV and 18 MV beams. BrainLab micromultileaf m3 collimators were used to generate 6×6 mm$^2$ to 100×100 mm$^2$ fields. A maximum variation of 11.7% was found in measured *OF*'s for the 6×6 mm$^2$ field size. Such variations in obtained *OF*'s indicate significant impact of dosimeter material, geometry and shape on *OF* measurements when the field sizes are comparable to the detector size.

To address the problems associated with small field dosimetry Alfonso *et al* (2008) proposed a formalism introducing a small field output correction factor ($k_{Q_{clin},Q_{msr}}^{f_{clin},f_{msr}}$), since $k_Q$ factors (Almond *et al* 1999) that were previously defined in the external beam radiotherapy code of practice are not sufficient to correct for detector response in small photon fields.

Current literature shows that deriving $k_{Q_{clin},Q_{msr}}^{f_{clin},f_{msr}}$ factors has been a considerable challenge for fields less than 10 mm diameter, and published results are rather contradictory (Tyler *et al* 2013, Dieterich and Sherouse 2011, Ralston *et al* 2012, 2014, Cranmer-Sargison *et al* 2012a, Benmakhlouf *et al* 2014, 2015). Bassinet *et al* (2013) measured the *OF* of small photon beams using several detectors and determined their values for $k_{Q_{clin},Q_{msr}}^{f_{clin},f_{msr}}$ factors. They found that EBT2 and microcubes have a close to unity correction factors. Ralston *et al* (2014) used EBT2 film to derive $k_{Q_{clin},Q_{msr}}^{f_{clin},f_{msr}}$ for the PTW (T60019) microDiamond detector and reported it over-responded by 4-5% in a 4-mm field. Underwood *et al* (2015) measured correction factors for microDiamond detector in 5 mm 6 MV field and reported an over-response of 4-5%. This agrees with recent IAEA-AAPM report on small field dosimetry (Palmans *et al* 2017). Other studies on the PTW microDiamond detector (Chalkley and Heyes 2014, Lárraga-Gutiérrez *et al* 2015, Morales *et al* 2014, Andreo *et al* 2016) contradict these results and concluded that microDiamond was almost equivalent to water for 6 MV fields as small as 5 mm diameter.

MC calculations of $k_{Q_{clin},Q_{msr}}^{f_{clin},f_{msr}}$ factors for very small (smaller than 4 mm diameter) fields are also not straightforward. In addition to the challenges associated with statistical uncertainties in dose calculations for very small volumes, there is also an ongoing debate in the literature (Marinelli *et al* 2016a, Andreo and Palmans 2016, Marinelli *et al* 2016b) on whether the lack of accurate design specifications for the microDiamond detector could be the source of inconsistency between experimental and MC derived $k_{Q_{clin},Q_{msr}}^{f_{clin},f_{msr}}$ factors. On one hand the size and shape of active volume is well established and its variations from detector to detector are very small (Marinelli *et al* 2016a). On the other hand, some metallic components within detector geometry that are seen in x-ray images, are not included in the diagrams provided by the manufacturer. This could potentially lead to discrepancies between experiment and MC results as indicated by Andreo *et al* (2016).

In this work, $k_{Q_{clin},Q_{msr}}^{f_{clin},f_{msr}}$ factors for microDiamond and unshielded EFD-3G detectors were derived experimentally (using GafChromic EBT3 films) and calculated using MC for a range of Varian TrueBeam 6 MV fields from 1.27 to 40 mm in diameter. $k_{Q_{clin},Q_{msr}}^{f_{clin},f_{msr}}$ factors were calculated for several detector orientations relative to the beam central axis (CAX). The impact of the possible variations in microDiamond detector inner schematics on the calculated $k_{Q_{clin},Q_{msr}}^{f_{clin},f_{msr}}$ correction factors was also investigated. This is the first report that investigates $k_{Q_{clin},Q_{msr}}^{f_{clin},f_{msr}}$ factors for fields smaller than 4 mm.

## 2. MATERIALS AND METHODS

### 2.1. Implementation of small field's dosimetry formalism for $k_{Q_{clin},Q_{msr}}^{f_{clin},f_{msr}}$ factor MC calculations and EBT3 film measurements.

Alfonso *et al* (2008), introduced dosimetry formalism for small and non-standard radiotherapy fields. A correction factor ($k_{Q_{clin},Q_{msr}}^{f_{clin},f_{msr}}$) was introduced to correct a detector response in small clinical fields ($f_{clin}$) relative to its response in the reference field ($f_{ref}$). For machines that cannot produce a 10×10 cm² reference field, a machine specific reference field ($f_{msr}$) is used instead. This output correction factor is defined as

$$k_{Q_{clin},Q_{msr}}^{f_{clin},f_{msr}} = \frac{D_{w,Q_{clin}}^{f_{clin}}/M_{Q_{clin}}^{f_{clin}}}{D_{w,Q_{msr}}^{f_{msr}}/M_{Q_{msr}}^{f_{msr}}} \quad (1)$$

where

$D_{w,Q_{clin}}^{f_{clin}}$ is absorbed dose to water at reference depth in a phantom for clinical field $f_{clin}$ of beam quality $Q_{clin}$.

$D_{w,Q_{msr}}^{f_{msr}}$ is absorbed dose to water at reference depth in a phantom for machine specific reference field $f_{msr}$ of beam quality $Q_{msr}$.

$M_{Q_{clin}}^{f_{clin}}$ is detector reading at reference depth in a phantom for clinical field $f_{clin}$ of beam quality $Q_{clin}$.

$M_{Q_{msr}}^{f_{msr}}$ is detector reading at reference depth in a phantom for machine specific reference field $f_{msr}$ of beam quality $Q_{msr}$.

The ratio $\frac{M_{Q_{clin}}^{f_{clin}}}{M_{Q_{msr}}^{f_{msr}}}$ in equation (1) corrects the detector response in the clinical field relative to the machine specific field and will be defined here as the detector output factor ($OF_{det}$).

$$OF_{det} = \frac{M_{Q_{clin}}^{f_{clin}}}{M_{Q_{msr}}^{f_{msr}}} \qquad (2)$$

In MC calculations for non-water equivalent detectors this factor is derived by modeling the detector geometry (and construction materials) and scoring the dose in the sensitive volume of the detector placed in clinical as well as reference fields.

The other ratio $\frac{D_{w,Q_{clin}}^{f_{clin}}}{D_{w,Q_{msr}}^{f_{msr}}}$ in equation (1) converts absorbed dose to water in the machine specific reference field to absorbed dose to water in a clinical field. It is commonly known as field output factor $\Omega_{Q_{clin},Q_{msr}}^{f_{clin},f_{msr}}$ and can be obtained using perturbation-free dosimeter (with $k_{Q_{clin},Q_{msr}}^{f_{clin},f_{msr}} = 1$, $\Omega_{Q_{clin},Q_{msr}}^{f_{clin},f_{msr}} = \frac{D_{w,Q_{clin}}^{f_{clin}}}{D_{w,Q_{msr}}^{f_{msr}}} = \frac{M_{Q_{clin}}^{f_{clin}}}{M_{Q_{msr}}^{f_{msr}}}$). In measurements with perturbation-free dosimeter $\Omega_{Q_{clin},Q_{msr}}^{f_{clin},f_{msr}} = OF_{det}$. In MC calculations, that are also perturbation-free in water, $\Omega_{Q_{clin},Q_{msr}}^{f_{clin},f_{msr}}$ is derived as the ratio of the doses from clinical and reference fields scored in small volumes of water.

In general, from equation (1), the field output factor $\Omega_{Q_{clin},Q_{msr}}^{f_{clin},f_{msr}}$ can be described as the detector output factor ($OF_{det}$) multiplied by its correction factor ($k_{Q_{clin},Q_{msr}}^{f_{clin},f_{msr}}$):

$$\Omega_{Q_{clin},Q_{msr}}^{f_{clin},f_{msr}} = OF_{det} \cdot k_{Q_{clin},Q_{msr}}^{f_{clin},f_{msr}} \qquad (3)$$

Previous studies showed that EBT2/3 films do not require correction factors in small field dosimetry (Novotny Josef et al 2009, Bassinet et al 2013, Lárraga-Gutiérrez 2014). Thus, in this work, EBT3 film is assumed to be the perturbation-free dosimeter ($k_{Q_{clin},Q_{msr}}^{f_{clin},f_{msr}} = 1$), and therefore $\Omega_{Q_{clin},Q_{msr}}^{f_{clin},f_{msr}} = OF_{EBT}$ was used in equation (3) to experimentally derive the $k_{Q_{clin},Q_{msr}}^{f_{clin},f_{msr}}$ factors for microDiamond and EFD-3G detectors as follows

$$k_{Q_{clin},Q_{msr}}^{f_{clin},f_{msr}} = \frac{OF_{det}}{\Omega_{Q_{clin},Q_{msr}}^{f_{clin},f_{msr}}} = \frac{OF_{det}}{OF_{EBT}} \qquad (4)$$

$k_{Q_{clin},Q_{msr}}^{f_{clin},f_{msr}}$ factor could also be looked at as a product of two independent corrections that arise from non-water equivalency and finite size of the detector sensitive volume:

$$k_{Q_{clin},Q_{msr}}^{f_{clin},f_{msr}} = k(_{Q_{clin},Q_{msr}}^{f_{clin},f_{msr}})_{no\,vol} \times P_{vol}^{det} \qquad (5)$$

Where, $k(_{Q_{clin},Q_{msr}}^{f_{clin},f_{msr}})_{no\,vol}$ is the correction factor for an imaginary detector with infinitely small sensitive volume but the same geometry and materials as the detector under study. Volume averaging correction factor for such an imaginary detector would be unity. Indeed, equation (5) includes an assumption that the difference in particle fluence through the real detector as compared to an imaginary one is negligible. Volume averaging correction factor $P_{vol}^{det}$ is well recognized as an essential correction in small field dosimetry, and has been described by Papaconstadopoulos et al (2014). $P_{vol}^{det}$ factor can be calculated using MC as the ratio of dose deposited in a very small volume of water ($D_w$) to the dose deposited in the volume of water ($D_{vol,w}$) equal to the detector sensitive volume

$$P_{vol}^{det} = \frac{D_w}{D_{vol,w}} \qquad (6)$$

Inclusion of these two factors into our modeling allowed investigation of the balance between energy response and the volume averaging in microDiamond and EFD-3G detectors.

## 2.2. Experimental measurements

Varian TrueBeam (Varian Medical Systems, Palo Alto, California) linac with BrainLab SRS cones (BrainLAB AG, Feldkirchen, Germany) was used to generate 6 MV circular fields. The BrainLab circular collimators were 10, 12.5, 15 and 40 mm nominal fields. Three in-house customized collimators were used to produce very small circular fields of 1.27, 2.46 and 3.77 mm in diameter. Measurements were taken with the detectors placed at the isocenter and 1.5 cm depth in Solid Water ("RMI-457", Gammex RMI, Middleton, WI). The relative output factors were obtained as in equation (3) from the ratio of the detector reading in clinical field to its reading in the reference field of 40 mm diameter.

PTW-60019 microDiamond (PTW-Freiburg, Germany) and IBA EFD-3G diode (IBA-Dosimetry, Germany) detectors were used to measure output factors. Both detectors were used with their stems perpendicular to the beam CAX. Thus, only $k_{Q_{clin},Q_{msr}}^{f_{clin},f_{msr}}$ factors for the perpendicular detector orientation were experimentally derived in this work. To make sure that the detectors were placed exactly at the center of the field, small shifts were introduced to the detectors until maximum signal was measured. The accuracy of the detector alignment in the beam was about 0.1 mm in these measurements. Three readings were recorded for each collimator and the results were averaged. At least two sets of measurements were performed at different times for each detector to evaluate the setup uncertainties.

GafChromic® EBT3 film (Ashland, Specialty Ingredients, NJ) pieces of 5×5 cm² were used to obtain the beam profiles and output factors. The exposed pieces of film were scanned at 200-500 dpi scanner resolution. Output factor readings were extracted from a 3×3 pixels size area (0.4×0.4 - 0.15×0.15 mm²) at the center of the field. Film calibration, scanning and image processing were carried out as described previously (Alhakeem *et al* 2015) unless stated otherwise. Lateral dose profiles and *OF*s were then extracted from the films.

Dosimetric field sizes defined by the full width at half maximum (FWHM) (Cranmer-Sargison *et al* 2013) were determined for all fields and these were used in $OF_{det}$ comparisons with MC calculations as per (Cranmer-Sargison *et al* 2013).

$k_{Q_{clin},Q_{msr}}^{f_{clin},f_{msr}}$ factors for microDiamond and EFD-3G detectors were obtained from equation (4) where field output factor was derived from EBT3 measurements: $\Omega_{Q_{clin},Q_{msr}}^{f_{clin},f_{msr}} = OF_{EBT}$.

## 2.3. Monte Carlo simulation

BEAMnrc/DOSXYZnrc Monte Carlo codes (Rogers *et al* 2009, Walters *et al* 2005) were used to calculate the field output factor $\Omega_{Q_{clin},Q_{msr}}^{f_{clin},f_{msr}}$ (ratio of absorbed doses) and dose profiles in a water phantom with 1×1×1 mm³ voxels for the 10-40 mm cones and 0.1×0.1×0.25 mm³ voxels for the smaller cones. The modeled parts of the linac, as shown in figure 1, included monitor chamber and mirror above the Varian phase-space file (PSF) to account for the back-scatter factor in the absolute dose calculations as described by Zavgorodni *et al.* (2014). The *egs_chamber* code (Wulff *et al* 2008) was used to model the microDiamond and the EFD-3G diode detectors, and to calculate the dose $M_{Q_{clin/msr}}^{f_{clin/msr}}$ deposited in their active volumes.

*2.3.1 Source simulation*

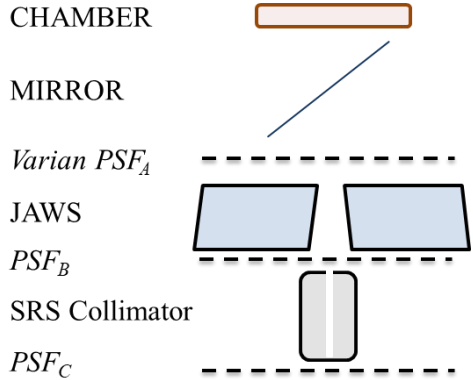

**Figure 1.** Schematic of the Monte Carlo model used in BEAMnrc calculations of the *OF*s. Shown are locations of Varian phase space (PSF$_A$) as well as the high-density particle phase-space (PSF$_B$) and small fields phase-space PSF$_C$ scored during the different stages of simulation.

The simulation was carried in two stages. First, using the BEAMnrc code forty Varian 6 MV photon phase-space files (PSF$_A$) were transported through the jaws, set at 5×5 cm$^2$, and ancillary PSFs were scored straight under the jaws. These forty ancillary files were then summed up into a single phase-space file (PSF$_B$) beneath the jaws as illustrated in figure 1. The resultant PSF$_B$ contained 200×10$^9$ particles. The large number of PSFs that was used to produce PSF$_B$ was essential for reduction of the latent variance as reported in our recent work (Alhakeem and Zavgorodni 2017). In the second stage, particles from the PSF$_B$ were propagated through a circular collimator and scored into another phase-space file (PSF$_C$) with a density of over 5 million particles per cm$^2$. Photon (PCUT) and electron (ECUT) cutoff energies were 0.01 MeV and 0.700 MeV, respectively.

These small field phase-space files (PSF$_C$), were used as a particle source for DOSXYZnrc and *egs_chamber* Monte Carlo codes to calculate lateral dose profiles, field output factors and $k_{Q_{clin},Q_{msr}}^{f_{clin},f_{msr}}$ factors.

*2.3.2. MC calculations to derive lateral profiles and field output factors*

PSF$_C$ files were used as a source in DOSXYZnrc to calculate lateral dose profiles as well as output factors from all cones in a water phantom. Photon and electron (PCUT, ECUT) cutoffs were 0.01 MeV and 0.521 MeV, respectively. The "Exact" boundary crossing algorithm was used along with condensed history electron step algorithm PRESTA-II. This provided sufficient accuracy of the dose deposition even for the smallest 0.1×0.1×0.25 mm$^3$ voxels used in this work.

Lateral dose profiles were obtained at a depth of 1.5 cm with source to surface distance (SSD) 98.5 cm and were benchmarked against EBT3 film measurements described in section 2.2. Field output factors for each collimator were calculated as the ratio of the doses scored in water for these collimators to the dose from the 40 mm diameter cone at 1.5 cm depth and SSD=98.5 cm.

For the three smallest cones attempts were made to obtain exact match of MC calculated profiles with profiles measured using film by varying the inner diameter of the collimator. However, due to the very time consuming process of collection and summation of ancillary PSFs required for calculating these profiles, exact match of the profiles was not feasible. The FWHM differences between calculated and measured profiles were in the range of 0.05 – 0.18 mm (Table 2), and were deemed acceptable as we compare $\Omega_{Q_{clin},Q_{msr}}^{f_{clin},f_{msr}}$ and $k_{Q_{clin},Q_{msr}}^{f_{clin},f_{msr}}$ against dosimetric rather than "nominal" field sizes. In table presentation (tables 3 -5) the values of MC calculated dosimetric field sizes were interpolated for direct comparison with measurement, in graphs the actual calculated values of dosimetric field sizes were used.

### 2.3.3. Deriving $k_{Q_{clin},Q_{msr}}^{f_{clin},f_{msr}}$ factor with MC calculations

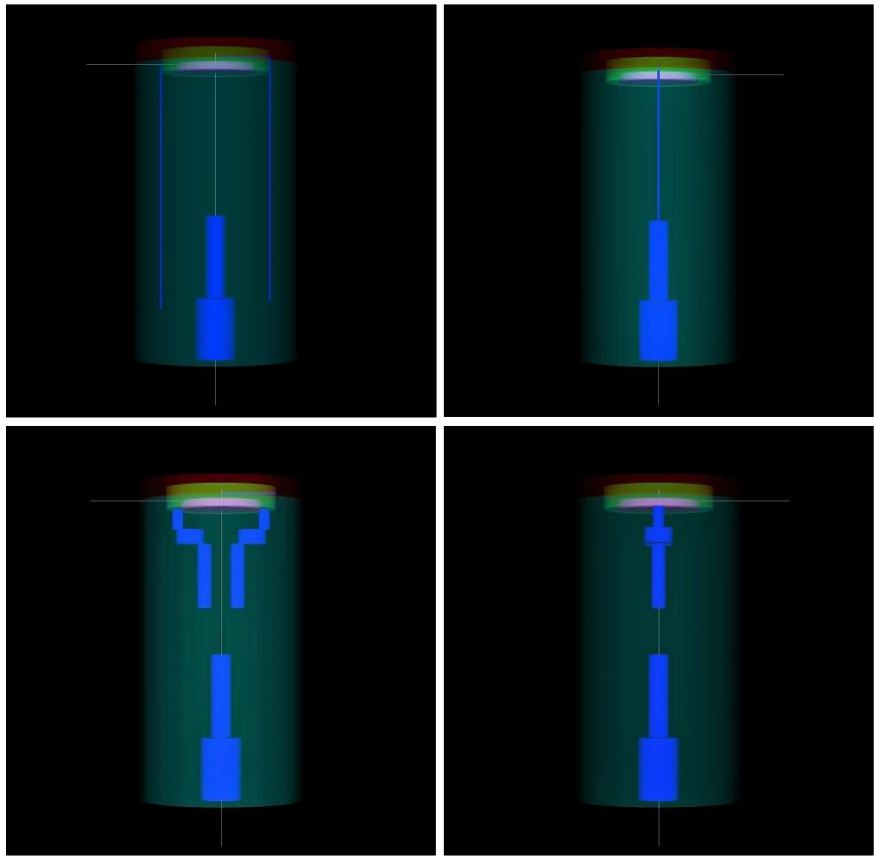

**Figure 2.** An *egs_view* (EGSnrc geometry viewing tool) image of two microDiamond detector models. On top, model (A) with no metallic connection pins, and bottom model (B) with metallic connection pins included.

**Table 1.** Detectors geometry and materials included in egs_chamber $k_{Q_{clin},Q_{msr}}^{f_{clin},f_{msr}}$ simulations.

| Detector | Active volume | | | | Capsule materials |
|---|---|---|---|---|---|
| | Material | Z$_{eff}$ | diameter mm | Thickness mm | |
| **microDiamond** | Carbon | 6 | 2.2 | 0.001 | RW3, epoxy resin, PMMA Aluminum |
| **EFD-3G diode** | Silicon | 14 | 2.0 | 0.06 | Epoxy resin |

MC code *egs_chamber* (Wulff *et al* 2008) was used to model two versions of the PTW-60019 microDiamond detector and EFD-3G detector.

Model A (figure 2) of PTW-60019 microDiamond detector was based on available manufacturer specifications and reported geometrical details (Pimpinella *et al* 2012, Ciancaglioni *et al* 2012, Mandapaka *et al* 2013). Model B includes extra metallic pins (Andreo *et al* (2016)) that were assumed to be made of aluminum. Other geometry and material specifications were exactly the same in these modeled versions. The material and geometry specifications for the microDiamond detector are summarized in table 1. *egs_view* (EGSnrc 3D geometry viewing tool) images for both models are presented in figure 2. $k_{Q_{clin},Q_{msr}}^{f_{clin},f_{msr}}$ factors were calculated for two detector orientations, parallel and perpendicular to the beam CAX as shown in figure 3.

The impact of detector rotation relative to its stem axis while in perpendicular orientation was also investigated for model B where potential impact of this rotation is expected to be greater. Therefore, for model B $k_{Q_{clin},Q_{msr}}^{f_{clin},f_{msr}}$ factors were calculated for the detector electrodes lined up along the beam CAX, (figure 3-a) and perpendicular to that (figure 3-b).

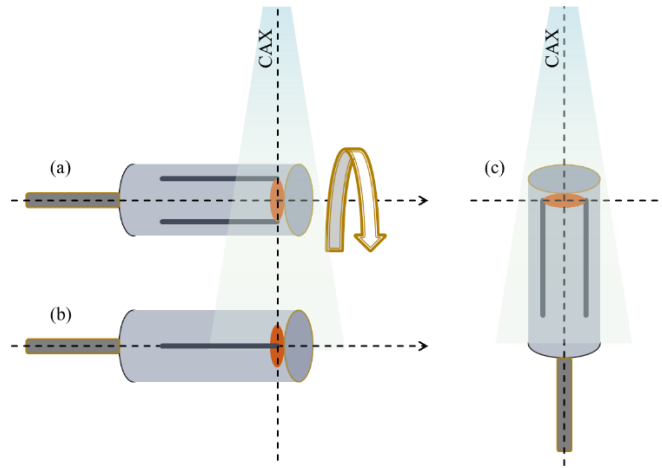

**Figure 3**. A diagram illustrating the three modeled orientations of the PTW microDiamond detector relative to the incident beam: a) detector's stem axis is perpendicular to beam CAX and electrodes are lined up along the beam axis, b) detector's stem axis is perpendicular to beam CAX, and electrodes are in a plane orthogonal to CAX, c) the detector's stem axis is aligned along the beam CAX.

For EFD-3G detector the model was simplified relative to its detailed geometry, and only included the silicon chip embedded in the epoxy housing as in Wang and Rogers (2007). Cranmer-Sargison et al (2012b) showed that such a simplified model is still accurate; for fields as small as 5x5 mm$^2$ it produced $OF_{det}$ s within 1% of those calculated by a complete model.

For each field size the $k_{Q_{clin},Q_{msr}}^{f_{clin},f_{msr}} = \frac{D_{w,Q_{clin}}^{f_{clin}}/M_{Q_{clin}}^{f_{clin}}}{D_{w,Q_{msr}}^{f_{msr}}/M_{Q_{msr}}^{f_{msr}}}$ factors were calculated for both microDiamond EFD-3G detectors. $D_{w,Q_{clin}}^{f_{clin}}$ and $D_{w,Q_{msr}}^{f_{msr}}$ were calculated in a very small cylindrical volume of water of 0.02 cm diameter and 0.03 cm thickness at 1.5 cm depth. $M_{Q_{clin}}^{f_{clin}}$ and $M_{Q_{msr}}^{f_{msr}}$ were the doses calculated in the detector active volume for the clinical and machine specific fields, respectively. Electron (ECUT) and photon (PCUT) cutoffs of 0.512 MeV and 0.01 MeV respectively, were applied. Cross-section enhancement (XCSE) factor of 128 was applied within the active volume and the surrounding layers.

*2.3.4. MC calculation of detector volume averaging factor*

Detector volume averaging factors $P_{vol}^{det}$ were calculated using equation (6). *egs_chamber* MC code was used to obtain the ratio of dose deposited in the detector sensitive volume replaced by water ($D_{vol,w}$) to the dose deposited in a very small cylindrical volume of water ($D_w$) of 0.02 cm diameter × 0.03 cm height located in the same position within the detector as the sensitive volume. $P_{vol}^{det}$ factors were calculated for the 1.27—10 mm collimators with detectors in both parallel and perpendicular orientations.

## 3. RESULTS

### 3.1. Benchmarking the Monte Carlo model

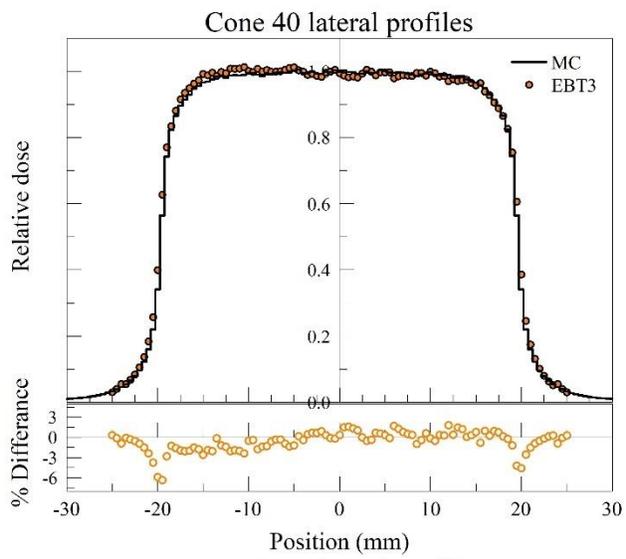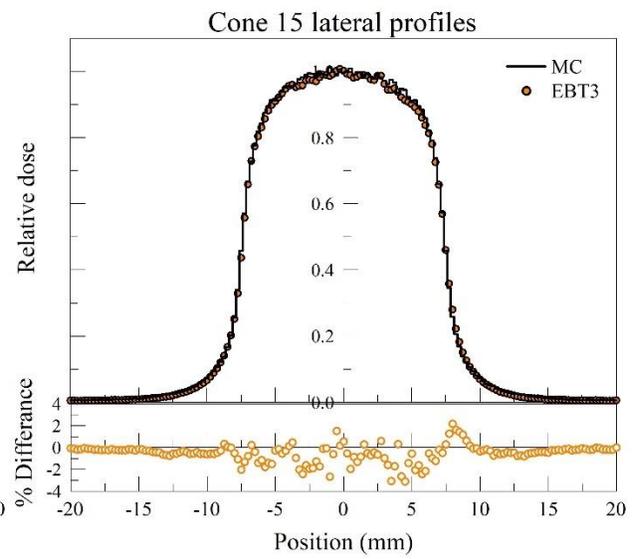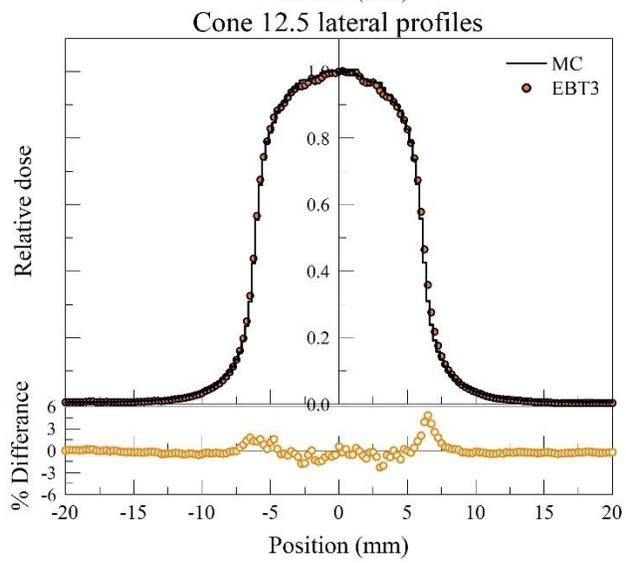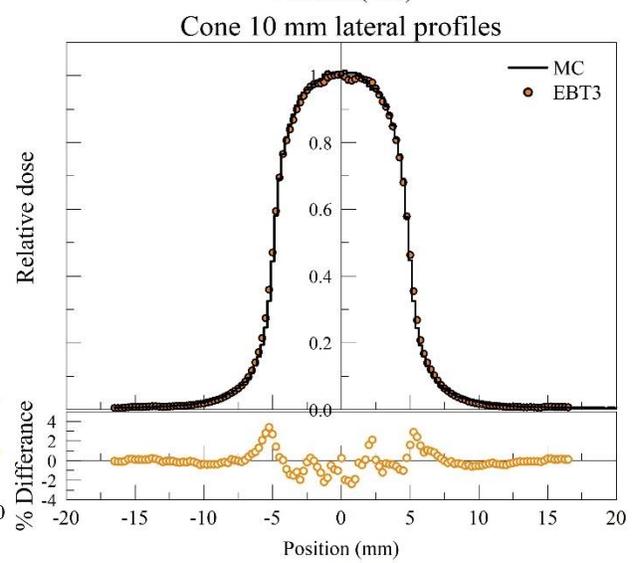

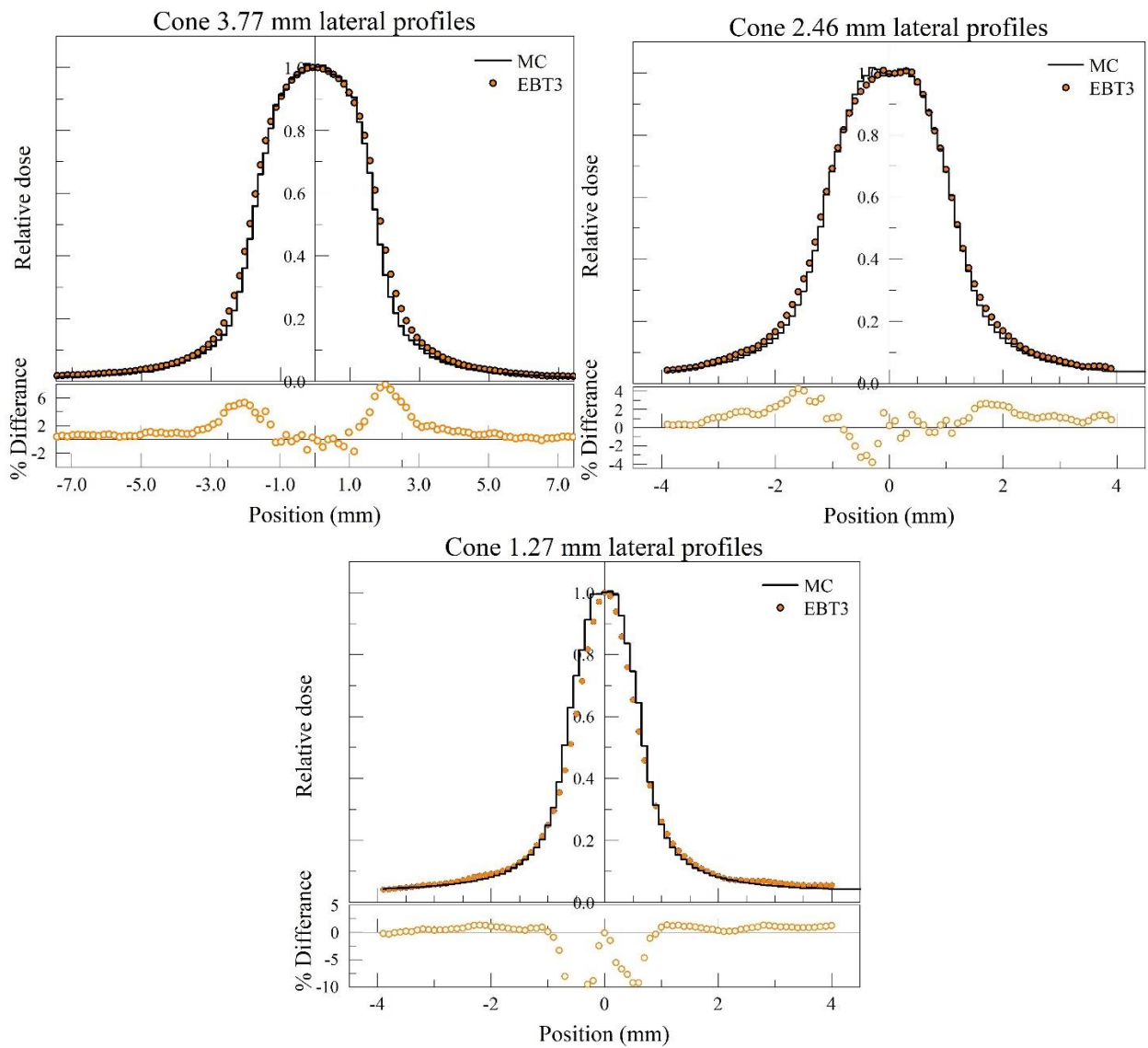

**Figure 4**. MC (DOSXYZnrc) calculated dose profiles for 1.27-40 mm collimators compared against EBT3 film measurements.

Figure 4 shows MC calculated profiles benchmarked against film measurements for 1.27 – 40 mm cones. Agreement inside the beam aperture between the two data sets is within 2.5% for all fields ≥10 mm. Mean distance-to-agreement (DTA) in the penumbra region (20-80%) for these cones was ≤ 0.2 mm. For the 1.27, 2.46 and 3.77 mm cones the mean DTA in the penumbra region was ≤ 0.15 mm.

### *3.2. Measured and calculated dosimetric fields*

**Table 2.** Dosimetric field sizes (FWHM) for the in-house collimators determined from EBT3 and MC profiles.

| $FWHM_{EBT3}$ (±σ) | $FWHM_{MC}$ (±σ) | %Relative difference |
| --- | --- | --- |

| 1.27 (±0.03) | 1.41 (±0.01) | 11% |
| 2.46 (±0.05) | 2.41 (±0.02) | -2.1% |
| 3.77 (±0.08) | 3.59 (±0.04) | -5% |

Table 2 shows the dosimetric field sizes for the three smallest collimators obtained from EBT3 films ($FWHM_{EBT3}$) and MC ($FWHM_{MC}$) lateral dose profiles. EBT3 film measurement based dosimetric field sizes are used to present the $OF_{det}$ s and $k_{Q_{clin},Q_{msr}}^{f_{clin},f_{msr}}$ factors in the following sections.

### 3.3. Detector output factors ($OF_{det}$)

**Table 3.** Detector output factors (relative to the 40 mm cone) were experimentally measured for 1.27 – 15 mm circular cones. PTW-60019 microDiamond and IBA EFD-3G diode detectors were used with their stems perpendicular to the beam CAX. Cone size represents the dosimetric field size derived from lateral dose profiles measured with EBT3 film. The bottom line in the table shows the largest difference between measured $OF_{det}$ and EBT3 measurement. Note that as MC calculations in water and EBT3 film measurement represent perturbation-free techniques their $k_{Q_{clin},Q_{msr}}^{f_{clin},f_{msr}} = 1$.

| Cone field size (mm) | 1.27† | 2.46† | 3.77† | 10 | 12.5 | 15 | 40 |
|---|---|---|---|---|---|---|---|
| microDiamond | 0.126 | 0.509 | 0.682 | 0.904 | 0.935 | 0.959 | 1 |
| EFD-3G | 0.119 | 0.504 | 0.648 | 0.884 | 0.914 | 0.943 | 1 |
| EBT3 | 0.145 | 0.500 | 0.626 | 0.883 | 0.927 | 0.944 | 1 |
| MC ($D_{w,Q_{clin}}^{f_{clin}}/D_{w,Q_{msr}}^{f_{msr}}$) | 0.148 | 0.489 | 0.616 | 0.873 | 0.928 | 0.945 | 1 |
| % Diff = $\frac{OF_{max}-OF_{EBT}}{OF_{EBT}} \times 100$ | -17.9 | 1.8 | 9.0 | 2.4 | -1.4 | 1.6 | - |

† In-house customized collimators

Table 3 shows the $OF_{det}$ obtained for the 1.27–15 mm cones. MC calculated output factors (in water) for the three smallest cones were linearly interpolated to match measured dosimetric fields of 1.27, 2.46 and 3.77 mm.

For the smallest 1.27 mm collimator, a discrepancy of -17.9% was found between EFD-3G detector and EBT3 film measurements. Maximum differences of 2.4%, 1.4% and 1.6% amongst the dosimeters were found for cones 10, 12.5 and 15 mm, respectively. EBT3 film obtained $OF_{EBT}$ agreed with the MC calculated output factors (dose to water ratio) within 1% for fields over 10 mm in diameter and within 2.2% for 3.77 mm and smaller fields.

### 3.4. $k_{Q_{clin},Q_{msr}}^{f_{clin},f_{msr}}$ correction factors for microDiamond and EFD-3G detectors

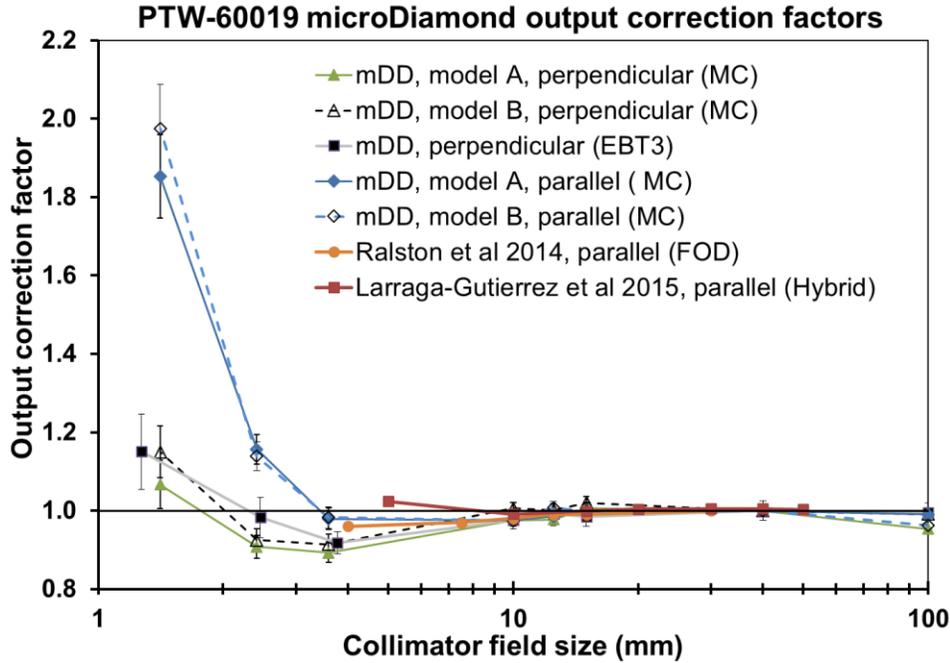

**Figure 5**. PTW-60019 microDiamond detector output correction factors $k_{Q_{clin},Q_{msr}}^{f_{clin},f_{msr}}$ for 1.27- 40 mm cones were determined experimentally (using the EBT3 film as the reference detector) as well as calculated with MC. MicroDiamond $k_{Q_{clin},Q_{msr}}^{f_{clin},f_{msr}}$ factors were calculated using two detector models, one with the absence (model A) and the other with the presence (model B) of metallic pins, and with two detector orientations. Previous studies by Ralston et al (2014) and Lárraga-Gutiérrez (2015) are added for comparison.

MC calculated and measured $k_{Q_{clin},Q_{msr}}^{f_{clin},f_{msr}}$ factors for the microDiamond detector are presented in figure 5. Correction factors for microDiamond detector in field sizes of ≥10 mm were small and ranged within 0.989—1.020 and within 0.975—1.010 for perpendicular and parallel orientations respectively. For 1.27, 2.46 and 3.77 mm cones, calculated $k_{Q_{clin},Q_{msr}}^{f_{clin},f_{msr}}$ factors for the microDiamond detector in perpendicular orientation were 41.75%, 18.8%, and 7.0%, respectively, smaller than those calculated for the detector in parallel orientation. The extra metallic pins included in model B had little effect on calculated $k_{Q_{clin},Q_{msr}}^{f_{clin},f_{msr}}$ factors for all field sizes except the smallest one. For the 1.27 mm field, model B $k_{Q_{clin},Q_{msr}}^{f_{clin},f_{msr}}$ factors for microDiamond detector in perpendicular and parallel orientations were ~7% higher than those for model A.

For all investigated fields the microDiamond detector, when setup in perpendicular orientation and rotated such that its electrodes lined up along the beam CAX, produced similar results to the setup with electrodes orthogonal to CAX. This indicates that any asymmetry of the detectors inner component relative to the incident fields had no significant impact on the results.

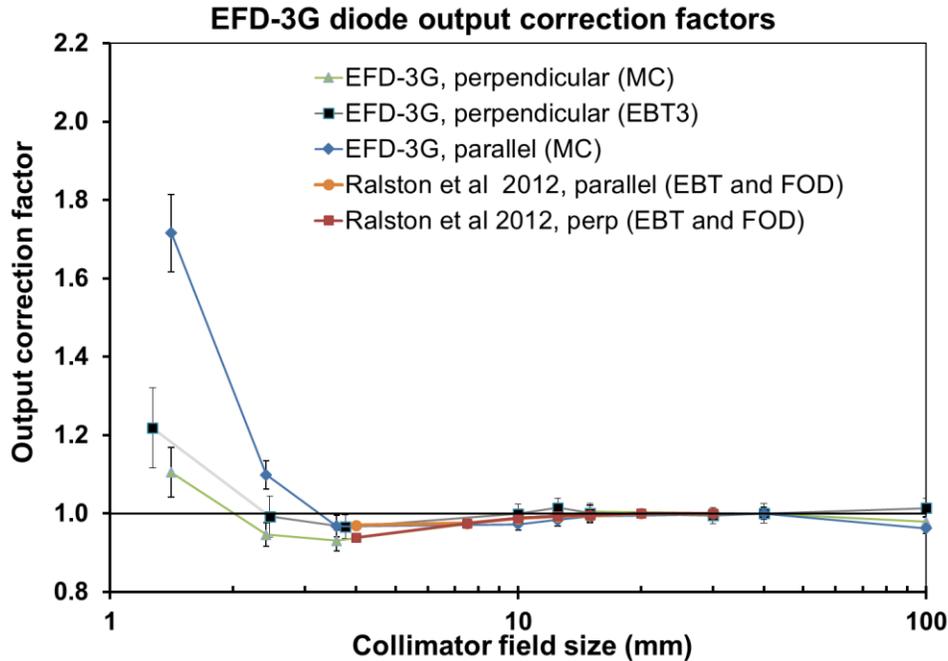

**Figure 6**. IBA EFD-3G unshielded diode detector output correction factors $k_{Q_{clin},Q_{msr}}^{f_{clin},f_{msr}}$ for 1.27- 40 mm cones were determined experimentally (using the EBT3 film as the reference detector) and calculated with MC . EFD-3G $k_{Q_{clin},Q_{msr}}^{f_{clin},f_{msr}}$ factors were calculated for two detector orientations and were measured using EBT3. Previous results by Ralston *et al* (2012) are added for comparison.

Figure 6 shows calculated and measured $k_{Q_{clin},Q_{msr}}^{f_{clin},f_{msr}}$ factors for the EFD-3G diode detector. The differences between parallel and perpendicular orientations in calculated $k_{Q_{clin},Q_{msr}}^{f_{clin},f_{msr}}$ factors were 35.6%, 13.9%, 3.8%, -1.6%, -1.1% and -1.3% for the 1.27, 2.46, 3.77, 12.5, and 15 mm cones, respectively.

Summary of the measured and calculated $k_{Q_{clin},Q_{msr}}^{f_{clin},f_{msr}}$ factors is shown in Table 4. Agreement between EBT3 and MC derived $k_{Q_{clin},Q_{msr}}^{f_{clin},f_{msr}}$ factors for microDiamond detector with its stem in the perpendicular orientation was within 3.7%% for 3.77 - 15 mm fields. Differences of 2.7% and 5.9% were found for the 1.27 and 2.46 mm fields, respectively. Agreement between EBT3 and MC derived $k_{Q_{clin},Q_{msr}}^{f_{clin},f_{msr}}$ factors for EFD-3G was less than 3.5% for the 3.77−15 mm fields. For the 1.27 and 2.46 mm fields, differences between measured and derived corrections were 7.5% and 4.7%, respectively.

**Table 4.** $k_{Q_{clin},Q_{msr}}^{f_{clin},f_{msr}}$ factors were measured and calculated for the PTW-60019 microDiamond and IBA EFD-3G unshielded diode detectors (both in perpendicular orientation) for a range of circular cones. Differences are shown between MC calculated and measured $k_{Q_{clin},Q_{msr}}^{f_{clin},f_{msr}}$ values for each detector.

| Collimator Field Size (mm) | $K_{Q_{clin},Q_{40mm}}^{f_{clin},f_{40mm}}$ | | | | *Relative Differences* | |
|---|---|---|---|---|---|---|
| | MC mDD | Measurement mDD | MC EFD | Measurement EFD | *% Diff<sub>mDD</sub>* | *% Diff<sub>EFD</sub>* |
| **1.27** | 1.182 | 1.151 | 1.127 | 1.219 | 2.7% | -7.5% |
| **2.46** | 0.924 | 0.982 | 0.945 | 0.992 | -5.9% | -4.7% |
| **3.77** | 0.912 | 0.918 | 0.932 | 0.966 | -0.7% | -3.5% |
| **10** | 0.989 | 0.977 | 0.989 | 0.999 | 1.2% | -1.0% |
| **12.5** | 1.001 | 0.991 | 0.995 | 1.014 | 1.0% | -1.9% |
| **15** | 1.020 | 0.984 | 1.005 | 1.001 | 3.7% | 0.4% |
| **40** | 1.000 | 1.000 | 1.000 | 1.000 | - | - |

### 3.5. Volume averaging

**Table 5.** $P_{vol}^{det} = D_w/D_{vol,w}$ factors calculated for the PTW-60019 microDiamond and IBA EFD-3G unshielded diode detectors in perpendicular and parallel orientations for 1.41-10 mm circular cones.

| | *Collimator field size (mm)* | *Volume averaging correction factor ($P_{vol}^{det}$)* | |
|---|---|---|---|
| | | *microDiamond* | *EFD-3G* |
| **Perpendicular stem orientation** | 1.41 | 1.390 | 1.316 |
| | 2.41 | 1.094 | 1.074 |
| | 3.59 | 1.014 | 1.007 |
| | 10 | 1.002 | 1.002 |
| **Parallel stem orientation** | 1.41 | 2.018 | 1.821 |
| | 2.41 | 1.237 | 1.179 |
| | 3.59 | 1.041 | 1.028 |
| | 10 | 1.001 | 1.000 |

Table 5 shows that volume averaging correction factors are considerably larger for detectors in the parallel orientation and for very small cones (1.41 – 3.59 mm). For microDiamond detector in the parallel orientation irradiated using 1.41 mm, 2.41 mm and the 3.59 mm collimators, $P_{vol}^{det}$ factors were 45.2%, 13.1% and 2.7% respectively higher than those with the detector perpendicularly orientated. Similarly, differences in $P_{vol}^{det}$ values for EFD-3G in parallel and perpendicular orientations were 38.4%, 9.8% and 2.1% for 1.41 mm, 2.41 mm and the 3.59 mm cones, respectively. MicroDiamond detector $P_{vol}^{det}$ factors were found to be larger than those for EFD-3G detector. This is because the microDiamond sensitive volume is larger than that of the EFD-3G detector.

### 3.6. Estimated uncertainties
#### 3.6.1. Estimated uncertainty in measurements

The uncertainties reported in $OF_{det}$ and the correction factors are a quadrature sum of type-A and type-B uncertainties using a one standard deviation confidence interval (1 σ) (BIPM et al 2008).

Type-A uncertainties were estimated from the standard deviation of three repeated measurements by microDiamond and EFD-3G detectors for each collimator. Type-A uncertainty in microDiamond and EFD-3G detectors measured $OF_{det}$ was estimated to be less than 0.6% for all fields. Type-B uncertainties due to setup and reproducibility were estimated for both detectors from the standard deviation of the results acquired from three setups assembled at different times. Type-B uncertainties were 1.27% for 10 – 40 mm collimators. For 1.27 – 3.77 mm fields, estimated type-B uncertainties were 5.4%–1.9%. Total uncertainty was derived through summation in quadrature of type-A and type-B uncertainties. Total $OF_{det}$ uncertainty for microDiamond and EFD-3G detectors was found to be in the range 5.5% – 2% for the 1.27 – 40 mm collimators.

Type-A uncertainties in $OF_{EBT}$ were estimated as described in Devic et al (2005) and were less than 1.4% for all fields. Type-B uncertainty was estimated considering only the setup and machine output variations and found to be less than 1.5% for fields greater than 10 mm. For the 1.27 – 3.77 mm fields, the estimated type-B uncertainties were 6.2% – 2%. This increase in the type-B uncertainty for small collimators was mainly due to the imperfect alignment of these collimators along the beam axis. EBT3 $OF_{EBT}$ combined uncertainty was 6.4% – 2% for the 1.27 – 40 mm fields.

Similarly, total uncertainty (type A and B) propagated into the $k_{Q_{clin},Q_{msr}}^{f_{clin},f_{msr}}$ estimate was 8% – 2.5% for the 1.27 – 40 mm fields.

*3.6.2. Estimated uncertainty in Monte Carlo calculations*

Latent variance (Sempau et al 2001) originating from the phase spaces provided by Varian for MC calculations has been estimated elsewhere (Alhakeem and Zavgorodni 2017). This sets the lower limit on achievable statistical uncertainties (type-A) in MC calculations, that in our simulations ranged within 3.5%–1.5% for 1.27 –3.77 mm collimators. Type-B uncertainties in MC calculations were estimated based on results reported by Francescon et al (2011). The overall uncertainties in the calculated field output factors $\Omega_{Q_{clin},Q_{msr}}^{f_{clin},f_{msr}}$ were 4%, 2.3%, and 1.5% for the 1.27 mm, 2.46 mm and all the other fields respectively. Total estimated uncertainties in the $k_{Q_{clin},Q_{msr}}^{f_{clin},f_{msr}}$ factors were 5.8%, 3.2%, 2.8% and less than 2% for 1.27 mm, 2.46 mm, 3.77 mm and all the other fields, respectively.

## 4. DISCUSSION

To the best of our knowledge, our study is the first to evaluate $k_{Q_{clin},Q_{msr}}^{f_{clin},f_{msr}}$ correction factors for microDiamond and EFD-3G (unshielded diode) detectors in very small fields of 1.27 – 3.77 mm diameter and demonstrates over and under-response of these detectors in such fields. In addition, we showed that the magnitude of correction is dependent on detector stem orientation with respect to the beam axis.

In this study $k_{Q_{clin},Q_{msr}}^{f_{clin},f_{msr}}$ factors were derived using MC for microDiamond detector in different orientations. Previous studies have only focused on investigating the detector response in parallel orientation, and many studies (Chalkley and Heyes 2014, Lárraga-Gutiérrez *et al* 2015, Morales *et al* 2014) showed that microDiamond detector $k_{Q_{clin},Q_{msr}}^{f_{clin},f_{msr}}$ factors were almost equal to unity for fields of over 4 mm diameter. For 4 – 5 mm fields microDiamond detector over-response by 4% – 5% was measured by Ralston *et al* 2014, Azangwe *et al* 2014, and Underwood *et al* 2015. Barrett and Knill (2016) reported a MC obtained $k_{Q_{clin},Q_{msr}}^{f_{clin},f_{msr}}$ factor of 0.969 (over-response of 3.1%) for microDiamond in a 4 mm field. Our MC results for a 3.77 mm field with the detector in the same (parallel) orientation demonstrated a smaller over-response of 1.8% ($k_{Q_{clin},Q_{msr}}^{f_{clin},f_{msr}}$=0.982). Therefore our MC calculated $k_{Q_{clin},Q_{msr}}^{f_{clin},f_{msr}}$ factor for a similar (3.77 mm) field agreed with results by Ralston *et al* 2014, Azangwe *et al* 2014, Underwood *et al* 2015, Barrett and Knill 2016 to within 1.3% - 3.2%.

While the orientation dependence of correction factors changes insignificantly for the fields of over 5 mm, for smaller fields this dependence is much more pronounced. For the field of 3.77 mm the microDiamond detector in parallel orientation required a considerably smaller correction (0.982) than for the perpendicular orientation (0.915). This is due to a larger volume averaging factor off-setting the detector over-response. For the 1.27 mm cone, the volume averaging contribution dominates in both detector orientations as demonstrated by the $k_{Q_{clin},Q_{msr}}^{f_{clin},f_{msr}}$ factors being larger than unity.

Francescon *et al* (2017), Coste *et al* (2017) and Andreo *et al* (2016) reported MC calculated $k_{Q_{clin},Q_{msr}}^{f_{clin},f_{msr}}$ factors for microDiamond in parallel orientation. Reported corrections from these studies (0.987-1.000) agreed with our calculations to within -1.6% to 2% for field sizes of 10 mm and larger. The smallest field they investigated was 5 mm, and the calculated $k_{Q_{clin},Q_{msr}}^{f_{clin},f_{msr}}$ factor was ~0.995-1.007. For comparison with the above references, our interpolated $k_{Q_{clin},Q_{msr}}^{f_{clin},f_{msr}}$ factor is ~0.98 for a 5 mm diameter field. Therefore for this field our results agree with Francescon *et al* and Andreo *et al* within uncertainty of our calculations (~ ± 2%). Our calculated microDiamond correction factors compared to results by other studies are shown in figure 7.

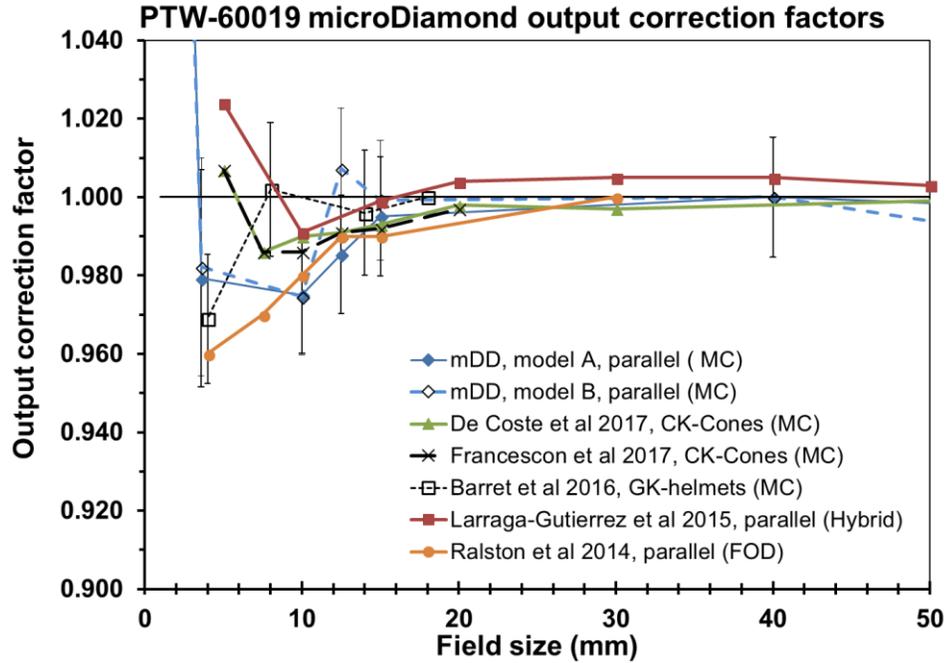

**Figure 7**. Comparison of MC obtained PTW-60019 microDiamond detector $k_{Q_{clin},Q_{msr}}^{f_{clin},f_{msr}}$ factors of figure 5 with results published by other studies.

Diode detector over-response has been previously investigated (Scott *et al* 2008, Francescon *et al* 2008, Ralston *et al* 2012, Bassinet *et al* 2013, Lechner *et al* 2013) and attributed mainly to higher atomic number and density of the diode's sensitive volume (silicon) compared to water. An in-depth interpretation of diode behavior in such fields was provided by Andreo *et al*. (2017), who attributed the over-response to the larger mean excitation energy (*I*-values) of silicon compared to that of water. Scott *et al* (2008) reported over-response of unshielded diode detectors, though its magnitude was lower compared to shielded ones. They reported an over-response of 4.5% by unshielded diode types in 5×5 mm$^2$ field. Bassinet *et al* (2013) measured small (4–15 mm diameter) fields' *OF*s, using different diode detectors (EDGE diode, PTW 60016, PTW 60017). For 4 mm and 10 mm cones measured *OF*s were higher than the mean *OF* measured by the EBT2 and LiF microcube detectors by 3–6% and 3.3–4.5% respectively. Again, this is consistent with our study, where *OF*$_{det}$ measured with EFD-3G diode detector for 3.77 mm filed size (in *perpendicular* orientation) was 3.5% and 5.0% higher than EBT3 film *OF*$_{EBT}$ and MC obtained dose to water ratios, respectively.

The calculated volume correction factors ($P_{vol}^{det}$) for both detectors were almost unity for fields greater than 10 mm similarly to previous reports (Papaconstadopoulos *et al* 2017, Ralston *et al* 2014, 2012). Ralston et al (2012) found that for the 4-mm cone the volume averaging correction of EFD-3G diode detector used in the parallel orientation was 1.9% higher than when it was perpendicular. Our MC calculated $P_{vol}^{det}$ factor for the 3.77 and 2.46 mm fields for the EFD-3G oriented in parallel were 2% and 9.8%, respectively, higher than the detector $P_{vol}^{det}$ factor in for a perpendicular orientation. Our MC calculated $P_{vol}^{det}$ factor for the microDiamond detector (parallel orientation) was 1.041 for the 3.77 mm cone and is consistent with $P_{vol}^{det}$ factor equal to 1.040 as measured by Ralston *et al* (2014) for the 4-mm cone.

This study showed that there were no significant differences between model A and model B of the microDiamond detector for all fields ≥ 2.46 mm. This means that including extra metallic connection pins into the detector model had no effect on the calculated results. In case of 1.27 mm field, $k_{Q_{clin},Q_{msr}}^{f_{clin},f_{msr}}$ for model B factor was ~7% higher than that for model A. This applies for both parallel and perpendicular detector stem orientations. A possible explanation for this result is that metallic pins included in model B acted as a shield preventing some of the lower energy particles from reaching the active volume and causing a lower detected signal (larger $k_{Q_{clin},Q_{msr}}^{f_{clin},f_{msr}}$ factor). However, the calculated values of $k_{Q_{clin},Q_{msr}}^{f_{clin},f_{msr}}$ factors are within the error bars and therefore this result is not definitive.

It is practically convenient to use the microDiamond detector in the perpendicular orientation for obtaining beam $OF_{det}$ in solid water phantoms, and this orientation also provides better spatial resolution. To our knowledge, this is the first paper to report experimental and MC obtained small field (1.27-15 mm diameter) $k_{Q_{clin},Q_{msr}}^{f_{clin},f_{msr}}$ correction factors for the microDiamond detector in a perpendicular orientation. For the three smallest fields (1.27 mm- 3.77 mm diameter) our MC and experimentally obtained $k_{Q_{clin},Q_{msr}}^{f_{clin},f_{msr}}$ factors were in agreement with each other to within 5.9% and 7.5% for the microDiamond and the EFD-3G diode detectors, respectively. For other field sizes, the MC calculated and EBT3 film measured $k_{Q_{clin},Q_{msr}}^{f_{clin},f_{msr}}$ factors were in agreement with each other to within 3.7% and 1.9% for the microDiamond and the EFD diode detectors respectively (table 4).

We can see three main possible reasons for the differences between experimental and MC derived correction factors. Firstly, the MC phase-space used in the calculations may not accurately represent the actual particle fluence for small fields. Secondly, geometry and materials of the sensitive volumes in MC calculations may not accurately represent those in the real detectors and finally, collimator-detector misalignment could have artificially reduced measured signals resulting in higher correction factors.

The first reason is unlikely to be valid because the agreement, within uncertainty, between EBT3 and MC obtained $OF_{det}$ (both measured and calculated in perturbation-free setting) indicates that the electron source parameters in Varian linac head model, that produced phase-spaces used in this work, were tuned with sufficient accuracy to represent small beams. This agreement also supports MC modeling as well as alignment of the collimators along the beam axis.

Regarding to modeling of detector geometries, our results showed that possible difference in the geometry of connection pins in microDiamond detector did not produce a measurable impact on output factors. Andreo *et al* (2016) indicated that the effective measurement volume in microDiamond detector could have been considerably smaller (0.6 mm) than the 2.2 mm diameter stated by the manufacturer in the specifications. Our data do not support such a considerable change in the effective detector size. We evaluated that a reduction of the detector diameter to 0.6 mm would have decreased volume averaging factors by 30% and 8% for 1.27 mm and 2.46 mm diameter collimators respectively and consequently would increase the measured output by the same magnitude. Such $OF_{det}$ change is outside of our estimated measurement uncertainties.

Our EFD-3G detector model did not include any high-density metallic components (as per Cranmer-Sargison *et al* (2012b)). According to the findings by Benmakhlouf *et al.* (2016) the particle fluence spectra is mostly perturbed by the high atomic number "extra-cameral" EFD-3G detector components. Based on his findings our calculated factors for the approximate model (with no metallic parts) are expected to be somewhat larger compared to those measured for a real detector. Our MC calculated $k_{Q_{clin},Q_{msr}}^{f_{clin},f_{msr}}$ factors for the approximate EFD-3G detector model agreed with EBT3 film measured $k_{Q_{clin},Q_{msr}}^{f_{clin},f_{msr}}$ factors within experimental uncertainty, although they were 7.5% - 3.5% smaller in 1.27 - 3.77 mm fields. Note also that our data were produced for the detector in the perpendicular orientation as compared to the parallel orientation in the study by Benmakhlouf *et al.* We expect that in perpendicular orientation the relative effect of extra-cameral components would be reduced.

The differences between MC and experimentally derived correction factors can probably be attributed to a possible small detector mis-alignment along CAX. The accuracy of the detector alignment in the beam was about 0.1 mm in our measurements, and the possible shift of effective measurement point off-axis by such distance could produce a signal reduction of the magnitude comparable to the observed differences between calculated and measured $OF_{det}$.

## 5. CONCLUSIONS

PTW-60019 microDiamond and IBA EFD-3G detectors performed well for the fields ≥10 mm. The $OF_{det}$ measured by these detectors agreed to within ~3.5% with field output factors values obtained using MC and EBT3 films. This indicates that required corrections are small for $OF$ measured by both detectors in fields ≥10 mm.

In 3.8 mm field the PTW-60019 microDiamond detector over-responded compared to MC calculations and EBT3 measurement. The magnitude of the detector over-response in the perpendicular orientation was greater than for a parallel orientation.

In the smallest (1.27 mm) field the microDiamond detector under-responded compared to MC calculations and EBT3 measurement due to the dominant role of volume averaging effect.

We found that including the metallic connection pins in the microDiamond detector model is not necessary and does not alter the values of $k_{Q_{clin},Q_{msr}}^{f_{clin},f_{msr}}$ factor in fields larger than 2.46 mm diameter.

There was no difference found in the calculated $k_{Q_{clin},Q_{msr}}^{f_{clin},f_{msr}}$ factors when the microDiamond detector was rotated around stem axis. This indicates that such detector rotation should have no impact on the $OF_{det}$ measurements in small fields.

This study showed that microDiamond and EFD-3G detectors can be used in very small (1.27-3.77 mm) fields once determined $k_{Q_{clin},Q_{msr}}^{f_{clin},f_{msr}}$ corrections are applied. Expected uncertainty of such measurements will be in the range of 8%-2.5%.

## 6. ACKNOWLEDGEMENTS

The authors would like to thank Mr. Steve Gray for precision manufacturing of the small collimators. We extend our thanks to Dr. Wayne Beckham for reading the manuscript and providing his valuable feedback.